\documentclass[twocolumn,eqsecnum,aps]{revtex4}
\usepackage[dvips]{epsfig,color}
\usepackage{graphicx}
\usepackage{amsmath,amsfonts,amsbsy,amssymb}
\usepackage{tabularx}

\begin{document}

\title{ 
Geometrical Construction of   Supertwistor Theory
}
\author{Kazuki Hasebe}
\affiliation{Department of General Education, Takuma National College of Technology,   Takuma-cho, Mitoyo-city, Kagawa 769-1192, Japan \\
Email: hasebe@dg.takuma-ct.ac.jp}

\begin{abstract}

Supertwistor theory is geometrically constructed based on the SUSY Hopf map. We  derive  a new  incidence relation for the geometrical  supertwistor theory.
The  present supertwistor  exhibits remarkable properties:   
  Minkowski space  need not  be   complexified to introduce spin degrees of freedom, and  even number SUSY is {automatically} incorporated 
 by the geometrical set-up.
We also develop a theory for  massless free particle in  Minkowski superspace, which physically corresponds to the geometrical supertwistor theory.  
The spin degrees of freedom are originated from fermionic momenta as well as fermionic coordinates.  
The geometrical supertwistor is quantized  to reproduce same physical contents as in the original supertwistor theory. 
Relationships  to superspin formalism and  SUSY quantum Hall effect  are also discussed.

\end{abstract}

\maketitle

\section{Introduction}

As is well known, twistor theory is an approach towards a  geometrical quantization of  space-time, originally proposed by Penrose \cite{TheCentralProgrammePenrose}.
In the twistor program, 
the twistor space is  regarded more fundamental 
than   space-time.
A light ray (massless particle) has  special importance  in  twistor theory,
and  
the twistor space is  naturally introduced as  parameter space of   massless particle.  
A time slice of light-cone is given by a celestial sphere, and  the  mathematical foundation  of the twistor theory is intimately related to  the Hopf map:
\begin{equation}
S^{3}\rightarrow S^2.
\end{equation}
This particular notion of the nontrivial homotopy  from sphere to sphere in different dimensions plays a crucial role in  constructing the  twistor theory \cite{hep-th/9310115}.
It is known that 2-dimensional sphere  is  a special manifold that accommodates  complex structure, and   mathematical progress initiated by the twistor formalism  
has exclusively indebted  to analytic properties of  the twistor space 
\cite{penroselmp1969,atiwarcmp55}. 

 In this paper, we construct a supersymmetric  extension of twistor theory (supertwistor theory) based on a purely geometrical set-up: the supersymmetric extension of the 
 Hopf map (SUSY Hopf map) \cite{J.Math.Phys.31(1990)45,math-ph/9907020,hep-th/0409230}: 
\begin{equation}
S^{3|2}\rightarrow S^{2|2}.
\end{equation}
The fermionic components are geometrically introduced by the SUSY Hopf map, and  bring  spin degrees of  freedom. 
In the conventional twistor theory, 
 the complexified Minkowski space is postulated  to introduce spin degrees of freedom, and the hermiticity of  Minkowski space is sacrificed.
In the supertwistor theory first introduced by Ferber \cite{FerberNucl.Phys.B132},
  imaginary coordinates of the complexified Minkowski space are replaced by fermion bispinor forms, but the complexified Minkowski space is still postulated. 
In the present geometrical approach,   Minkowski space need $\it{not}$  be complexified, and the  hermiticity of  Minkowski space is 
 promoted to the super-hermiticity in  superspace.
The incidence relation is  also naturally promoted to a SUSY framework, and  provides  a new  
 nonlocal relation between   Minkowski
  superspace and  supertwistor space. 
The supersymmetry  has a geometrical 
meaning given by  the SUSY Hopf map, and the number of supersymmetry  always takes  even number.
 It is known that  the number of supersymmetry has to be  even  to provide  integer or half integer helicity multiplets in  quantized supertwistor theory \cite{ptp7018}, and the  number of supersymmetry  
has been conventionally fixed {by} {hands}.  In the present approach, even number of supersymmetry is $\it{automatically}$ incorporated   by the geometrical set-up.

This paper is organized as follows.
In Sect.\ref{bosonhopfmaptwistor}, we review the Hopf map and its relation to  the twistor theory.
In Sect.\ref{sectsusyhopfandsuperspacematrix}, replacing the Hopf map with the SUSY Hopf map, 
 we develop a geometrical  supertwistor theory.
Properties of the  super incidence relation with emphasis on  differences to  Ferber's approach  are discussed in  Sect.\ref{superincesec}.
In Sect.\ref{supertwistorsection}, 
  we explore a   massless particle model in Minkowski superspace, which corresponds to the geometrical  supertwistor formalism. In Sect.\ref{secsupertwistorquantiz},   the  geometrical supertwistor  is quantized to  yield  same physical contents  obtained in  the original supertwistor theory. Relations to Bloch supersphere and  SUSY quantum Hall effect are discussed 
in Sect.\ref{relationstoquantumapin}. 
Sect.\ref{summarysection} is devoted to  summary and discussion.
In Appen.\ref{defofsumatapp},  several  definitions used in  super Lie group  are briefly explained.

\section{Review of Hopf Map and Incidence Relation}\label{bosonhopfmaptwistor}

First, we introduce  Hopf map and discuss its relation to the twistor theory.
The Hopf map $S^3\rightarrow S^2$ is explicitly given by 
\begin{equation}
\phi \rightarrow x^a=\phi^{\dagger}\sigma^a\phi,
\label{explHopf}
\end{equation}
where $\phi$ is a normalized two-component complex (Hopf) spinor: $\phi^{\dagger}\phi=1$, and $\sigma^a$ $(a=1,2,3)$ denote the Pauli matrices. 
By the normalization constraint, $\phi$ is regarded as the coordinates on $S^3$, and  $x^a$ defined by (\ref{explHopf}) satisfy the relation ${x^a}x^a=1$ that represents two-sphere with unit radius.
The Hopf map is the  template for more complicated  twistor theory, and as a preparation, we exploit its  basic features here. By reversing the Hopf map (\ref{explHopf}), the Hopf spinor is  given by 
\begin{equation}
\phi=
\begin{pmatrix}
\phi_1\\
\phi_2
\end{pmatrix}
=\frac{1}{\sqrt{2(1+x_3)}}
\begin{pmatrix}
1+x^3\\
x^1+ix^2
\end{pmatrix}\cdot e^{i\chi},
\label{explihopfspi}
\end{equation}
where $e^{i\chi}$ is the $U(1)$ phase factor canceled in the mapping (\ref{explHopf}), and the projective Hopf spinor space is 
defined as $S^3/S^1\approx S^2\approx {{\mathbb{C}}}{P}^1$. 
The Hopf map (\ref{explHopf}) suggests that  the Hopf spinor is a zero-mode of the ``space-matrix'' $r$: 
\begin{equation}
r=-1+x^a\sigma^a=
\begin{pmatrix}
-1+x^3 & x^1-ix^2 \\
x^1+ix^2 & -1 -x^3
\end{pmatrix}.
\end{equation}
With the stereographic coordinates  $x=x^1/(1+x^3)$ and $y=x^2/(1+x^3)$, the Hopf spinor is rewritten as   
\begin{equation}
\phi=
\begin{pmatrix}
\phi_1\\
\phi_2
\end{pmatrix}=
\frac{1}{\sqrt{1+x^2+y^2}}
\begin{pmatrix}
1 \\
x+iy
\end{pmatrix}\cdot e^{i\chi},
\end{equation}
and the upper component and  lower component in the Hopf spinor  is simply related as 
\begin{equation}
\phi_1=(x+iy)\phi_2. \label{simplestinci}
\end{equation}
(\ref{simplestinci}) is the simplest  incident relation that specifies 
one-to-one correspondence  between 
 points on the projective  Hopf spinor space $S^2$ and  points on the stereographic space  $R^2$ (except for the infinite distance). 
The incidence relation is gauge independent in the sense that the $U(1)$ phase factor does not appear in (\ref{simplestinci}). It is straightforward to generalize the above set-up for two-sphere with arbitrary radius $t$: 
\begin{equation}
t^2={x^a}x^a.\label{spherecond}
\end{equation}
The Hopf mapping is  rephrased as  
\begin{equation}
t=\phi^{\dagger}\phi,~~~x^a=\psi^{\dagger}\sigma^a\phi,
\label{4hopfmap}
\end{equation}
and the space  matrix is naturally promoted  to the ``space-time'' matrix $x$: 
\begin{equation}
x=-t+x^a\sigma^{a}=\begin{pmatrix}
-t+x^3 & x^1-ix^2 \\
x^1+ix^2 & -t -x^3
\end{pmatrix}.\label{spacetimematriori}
\end{equation}
It is important to notice if we identify  $t$ as time, the present two-sphere is   
regarded as time-slice of a light cone, i.e.  celestial sphere made of  light rays passing through the origin of Minkowski space. 
With this identification, the sphere condition (\ref{spherecond}) becomes  null vector condition for  $x^{\mu}=(t,x^1,x^2,x^3)$:
\begin{equation}
\eta_{\mu\nu}x^{\mu}x^{\nu}=\text{det}( x)
=0,
\end{equation}
where $\eta_{\mu\nu}=diag(-1,1,1,1)$.
The coordinates $x^{\mu}$ can be inversely obtained from $x$ as  
\begin{equation}
x^{\mu}=\eta_{\mu\nu}\text{tr}(x\sigma^\nu),
\end{equation}
where $\sigma^{\mu}=(1,\sigma^a)$. 
With the space-time matrix (\ref{spacetimematriori}),  the incidence relation in the twistor theory is given by 
\cite{TheCentralProgrammePenrose}  
\begin{equation}
\begin{pmatrix}
Z^1\\
Z^2
\end{pmatrix}
=i
\begin{pmatrix}
-t+x^3 & x^1-ix^2 \\
x^1+ix^2 & -t -x^3
\end{pmatrix}
\begin{pmatrix}
Z^3\\
Z^4
\end{pmatrix},\label{twistorince}
\end{equation}
where $Z^a=(Z^1,Z^2,Z^3,Z^4)$ represent  twistor variables, and $x^{\mu}$ are real coordinates in Minkowski space (and are not necessarily a null vector). As in the simplest incidence relation (\ref{simplestinci}), (\ref{twistorince}) specifies  relations between 
 points in the twistor space and space-time events in  Minkowski space. 
Conventionally, two spinor components  of the twistor are introduced  $Z^a=(\omega^{\alpha},\pi_{\beta})$ ($\alpha,\beta=1,2$), and   the incidence relation is written as 
\begin{equation}
\omega^{\alpha}=ix^{\alpha\beta}\pi_{\beta}.
\label{twistorince2}
\end{equation}
The space-time matrix is not affected by any complex scaling of the twistor variables, and the projective twistor space is defined as $\mathbb{C}P^3=S^7/S^1$ where $S^1$  represents the overall $U(1)$ phase  freedom.
Unlike the simplest version (\ref{simplestinci}), the incidence relation (\ref{twistorince2}) connects the space-time events and the twistor points nonlocally.
When a point in  twistor space is given, the corresponding space-time point is determined  up to the gauge transformation 
\begin{equation}
x^{\alpha\beta}\rightarrow x^{\alpha\beta}+a {\pi^{\alpha}}^*\pi^{\beta},
\end{equation}
where $a$ is an arbitrary real parameter to keep the hermiticity of $x^{\alpha\beta}$, and $\pi^{\alpha}$ is defined as  $\pi^{\alpha}=(-\pi_2,\pi_1)$. 
Such gauge degree of freedom corresponds to a null direction  in Minkowski space, since the null vector $p^{\mu}$ is constructed by the gauge part as 
\begin{equation}
p^{\mu}=-2\eta_{\mu\nu}(\sigma^{\nu})_{\beta}^{~~\alpha}{\pi^{\alpha}}^*\pi^{\beta}.
\end{equation}
 Thus, a point in the twistor space is nonlocally  
tranformed to a light ray in  Minkowski space. %
The inverse tranformation from Minkowski space to the twistor space
 is explanied as follows. 
Here, the coordinates in Minkowski space  are supposed to be  real,  then  the twistors  satisfy the null condition:
\begin{equation}
Z_a^*Z^a=0,
\label{nulltwiscond}
\end{equation}
where $Z^*_a=(\pi_{\alpha}^*,{\omega^{\alpha}}^*)$ represents the dual twistor.
Thus, the corresponding  (projective) twistor space is given by  the real five dimensional manifold called the projective null twistor space $\mathcal{PN}$. 
Provided the lower spinor component  $\pi_{\alpha}$ given, the entire twistor point is uniquely determined by the incidence relation. Since the lower component $\pi_{\alpha}$ geometrically represents  $S^2$, a point in  Minkowski space  corresponds   to  a two-sphere in the projective twistor space.  Such nonlocal transformations are the most particular feature in the twistor theory.

\section{SUSY Hopf Map and super space-time matrix}\label{sectsusyhopfandsuperspacematrix}

It has been reported the existence of the SUSY extension of the Hopf map \cite{J.Math.Phys.31(1990)45, math-ph/9907020,hep-th/0409230}, that is   the SUSY Hopf map: $S^{3|2}\rightarrow S^{2|2}$. 
The 3-component super (Hopf) spinor 
$\psi=(\psi_1,\psi_2,\psi_0)^t$, in which  $\psi_1$ and $\psi_2$ are Grassmann even components and $\psi_0$ is a  Grassmann odd component, plays a crucial role in constructing the SUSY Hopf map explicitly.  
The super Hopf spinor is normalized  as $\psi^{\ddagger}\psi=1$ with  $\psi^{\ddagger}=(\psi_1^*,\psi_2^*,-\psi_0^*)$. Here,  $*$ is not  the conventional 
complex conjugation but  the super-conjugation. (For the  definition of the super-conjudation, see Appen. \ref{defofsumatapp}.)
The SUSY Hopf map is  given by  
\begin{equation}
2\psi^{\ddagger}l^a\psi=x^{a},~~~2\psi^{\ddagger}l^{\alpha}\psi=\theta^{\alpha},
\label{psiladagger} 
\end{equation}
where $l^a$ and $l^{\alpha}$ are 
\begin{equation}
l^a =\frac{1}{2}
\begin{pmatrix}
\sigma^a & 0 \\
0 & 0 
\end{pmatrix}\!,~
l^{\alpha}=\frac{1}{2}
\begin{pmatrix}
 0 & \tau^{\alpha}\\
-(C\tau^{\alpha})^t & 0 
\end{pmatrix},
\end{equation}
with $\tau^1=(1,0)^t$, $\tau^2=(0,1)^t$, and $C$ is the charge conjugation matrix:  
\begin{equation}
 C=C_{\alpha\beta}=
\begin{pmatrix}
0 & 1 \\
-1 & 0 
\end{pmatrix},~~
C^t=C^{\alpha\beta}=
\begin{pmatrix}
0 & -1 \\
1 & 0 
\end{pmatrix}.\label{charconjmat}
\end{equation}
The spinor index is raised or lowered as  
 $\phi^{\alpha}=C^{\alpha\beta}\phi_{\beta}$ and   $\phi_{\alpha}=C_{\alpha\beta}\phi^{\beta}$.
$l^a$ and $l^{\alpha}$ satisfy the $OSp(1|2)$ algebra 
\begin{equation}
[l^a\!,l^b]\!=\! i\epsilon^{abc}l^c\!,~~[l^a\!,l^{\alpha}]\!=\!\frac{1}{2}(\sigma^a)_{\beta}^{~~\alpha}l^{\beta}\!,~~
\{l^{\alpha}\!,l^{\beta}\}\!=\!\frac{1}{2}(C\sigma^a)^{\alpha\beta}l^a\!,
\end{equation}
with $\epsilon^{123}=1$. 
Under the definition of the super-conjugation, $x^a$ and $\theta^{\alpha}$  (\ref{psiladagger})  become (pseudo-)real in the sense: 
\begin{equation}
{x^a}^*=x^a,~~~{\theta^{\alpha}}^*=\theta_{\alpha}, 
\end{equation}
where $\theta_{\alpha}=C_{\alpha\beta}\theta^{\alpha}$.
Besides, from the normalized super spinor $\psi$, $x_a$ and $\theta_{\alpha}$  satisfy the  condition 
\begin{equation}
x^a x^a+C_{\alpha\beta}\theta^{\alpha}\theta^{\beta}=1,
\end{equation}
which defines the supersphere with unit radius. Reversing (\ref{psiladagger}), the super Hopf spinor is expressed as  
\begin{equation}
\psi=\!\!
\begin{pmatrix}
\psi_1\\
\psi_2\\
\psi_0
\end{pmatrix}\!\!
=\!\!
\frac{1}{\sqrt{2(1+x^3)}}\!\!
\begin{pmatrix}
(1+x^3)(1-\frac{1}{4(1+x^3)}\theta C\theta)\\
(x^1+ix^2)(1+\frac{1}{4(1+x^3)}\theta C\theta)\\
(1+x^3)\theta^1+(x^1+ix^2)\theta^2
\end{pmatrix}\!\cdot  e^{i\chi}.
\label{explicitsuperhopf}
\end{equation}
Following the discussion in Sect.\ref{bosonhopfmaptwistor},   the  ``super space'' matrix 
 is similarly introduced as  
\begin{equation}
R=-2l^0+2x^{a}l^{a}+2C_{\alpha\beta}\theta^{\alpha} l^{\beta}
 =
\begin{pmatrix}
-1+x^3 & x^1-ix^2 & -\theta_2\\
x^1+ix^2 & -1-x^3 & \theta_1 \\
-\theta_1 & -\theta_2 & 1
\end{pmatrix},
\end{equation}
where $l^0$ is   
\begin{equation}
l^0=\frac{1}{2}
\begin{pmatrix}
1 & 0 & 0 \\
0 & 1 & 0 \\
0 & 0 & -1
\end{pmatrix}.
\end{equation}
The SUSY Hopf map (\ref{psiladagger}) suggests that  the super Hopf spinor is a zero-mode of the super space matrix: $R\psi=0$. The super stereographic coordinates are introduced as 
\begin{equation}
z= \frac{\psi_2}{\psi_1}=\frac{x^1+ix^2}{1+x^3}\biggl(1+\frac{1}{2(1+x^3)}
\theta C\theta\biggr),~~~
\theta= \frac{\psi_0}{\psi_1}=\theta^1+z\theta^2,
\end{equation}
and the SUSY Hopf spinor is represented as 
\begin{equation}
\psi=
\begin{pmatrix}
\psi_1\\
\psi_2\\
\psi_0
\end{pmatrix}
=
\frac{1}{\sqrt{1+zz^*+\theta \theta^*}}
\begin{pmatrix}
1\\
z\\
\theta
\end{pmatrix}\cdot e^{i\chi}.
\end{equation}
The super incidence relations are    
\begin{equation}
\psi_2=z \psi_1,~~~\psi_0= \theta \psi_1.
\end{equation}
It is easy to generalize the above discussion for the supersphere with  arbitrary radius $t$: 
\begin{equation}
t^2=x^ax^a+C_{\alpha\beta}\theta^{\alpha}\theta^{\beta}.
\label{supersphereondit}
\end{equation}
The corresponding super Hopf map is  given by 
\begin{equation}
2\psi^{\ddagger}\psi=t,~~~2\psi^{\ddagger}l^{a}\psi=x^{a},~~~2\psi^{\ddagger}l^{\alpha}\psi=\theta^{\alpha}.
\end{equation}
Identifying $t$ as time,  the present supersphere is regarded as a celestial supersphere that is equal to  a time slice of  super light-cone passing through the origin of Minkowski superspace.
Here, Minkowski  superspace is referred to $\mathcal{M}_{4|2}$ which has 6 (pseudo-)real coordinates,  4 of which are bosonic $x^{\mu}$ (${x^{\mu}}^*=x^{\mu}$), 2 are fermionic $\theta^{\alpha}$ (${\theta^{\alpha}}^*=\theta_{\alpha}$).  
The supersphere condition (\ref{supersphereondit}) is rephrased as the null super vector condition of  $x^{\mu}$ and $\theta^{\alpha}$: 
\begin{equation}
\eta_{\mu\nu}x^{\mu}x^{\nu}+C_{\alpha\beta}\theta^{\alpha}\theta^{\beta}
=-t \cdot {\text{sdet}} X=0,\label{supernullori}
\end{equation}
where the ``super space-time'' matrix $X$ is defined  as 
\begin{equation}
X=2\eta_{\mu\nu}x^{\mu}{l}^{\nu}+2C_{\alpha\beta}\theta^{\alpha}{l}^{\beta}=
\begin{pmatrix}
-t +x^3 & x^1-ix^2 & -\theta^2 \\
x^1+ix^2        &  -t-x^3 & \theta^1 \\
-\theta^1 & -\theta^2 & t 
\end{pmatrix}.
\label{explisupermatrix}
\end{equation}
The coordinates in Minkowski superspace are inversely obtained as 
\begin{equation}
x^0=-\text{str}(Xl^0),~~ x^{a}=\frac{1}{2}\eta_{\mu\nu}\text{str}(X l^{a} ),~~\theta^{\alpha}=\frac{1}{2}\text{str}(X l^{\alpha}).
\end{equation}
It should be noted that the super space-time matrix is super-hermitian under the definition of the super-adjoint $\ddagger$ in  Appen. \ref{defofsumatapp}:
\begin{equation}
{X}^{\ddagger}={X}.
\label{superherx}
\end{equation}

\section{ Super incidence relation}\label{superincesec}

Based on the analogy to the original incidence relation (\ref{twistorince}),  we introduce 
 the super incidence relation   as 
\begin{equation}
\begin{pmatrix}
Z^1\\
Z^2\\
\xi^1
\end{pmatrix}=i
\begin{pmatrix}
-t +x^3 & x^1-ix^2 & -\theta^2 \\
x^1+ix^2        &  -t-x^3 & \theta^1 \\
-\theta^1 & -\theta^2 & t 
\end{pmatrix} 
\begin{pmatrix}
Z^4\\
Z^5\\
\xi^2
\end{pmatrix},
\label{superincimatr}
\end{equation}
where $x^{\mu}$ and $\theta^{\alpha}$ need not be a super null vector (\ref{supernullori}).
Since the super space-time matrix is given by  the 3$\times$3  supermatrix,
 the corresponding   supertwistor has 6 components:  
$Z^A=(Z^1,Z^2,Z^3,Z^4,\xi^1,\xi^2)$ where 
$Z^1,Z^2,Z^3$ and $Z^4$ are Grassmann even while $\xi^1$ and $\xi^2$ are Grassmann odd quantities. It should be noted in the present approach, the number of the 
Grassmann odd components is fixed to 2 by the geometrical set-up.
In (\ref{superincimatr}), the super space-time matrix is invariant under the arbitrary complex scaling of the supertwistors, and the projective supertwistor space is 
defined by the projection of  the complex scaling, and hence has   the (pseudo-)real dimension $6|4$.
Introducing two  super spinors $\pi_A$ and $\omega^A$ 
\begin{align}
&\omega^A=(\omega^{\alpha},\omega)=(Z^1,Z^2,\xi^1),
 \nonumber\\
&\pi_A=(\pi_{\alpha},\pi)=(Z^3,Z^4,\xi^2),
\end{align}
the super incidence relation (\ref{superincimatr}) is written as 
\begin{align}
&\omega^{\alpha}=ix^{\alpha\beta}\pi_{\beta}-i\theta_{\alpha}\pi,\nonumber\\
&\omega=-i\theta^{\alpha}\pi_{\alpha}+it\pi.\label{spinorsuperinc}
\end{align}
The super incidence relation specifies nonlocal relations between 
 supertwistor space and  Minkowski superspace. With given  a point in twistor space, the corresponding point in Minkowski superspace cannot be determined uniquely due to the existence of  the gauge degree of freedom in (\ref{spinorsuperinc}):  
\begin{align}
&x^{\alpha\beta}\rightarrow x^{\alpha\beta}+a (2{\pi^{\alpha}}^*\pi^{\beta}-\delta^{\alpha\beta}\pi^*\pi),\nonumber\\
&\theta^{\alpha}\rightarrow \theta^{\alpha}-{a}({\pi_{\alpha}}^*\pi+\pi^*\pi^{\alpha}),\label{gaugesuperincidenceinx}
\end{align}
 where $a$ is an arbitrary real parameter.
The transformation of $t=x^{33}$ follows from that of $x^{\alpha\beta}$:
\begin{equation}
t\rightarrow t-{a}({\pi^1}^*\pi^1+{\pi^2}^*\pi^2-\pi^*\pi),
\end{equation}
and, similarly, $\theta_{\alpha}$ follows from  $\theta^{\alpha}$: 
\begin{equation}
\theta_{\alpha}\rightarrow \theta_{\alpha}+{a}({\pi^{\alpha}}^*\pi-\pi^*\pi_{\alpha}).
\end{equation}
Such gauge degrees of  freedom  represents a direction  of a super light ray (this will be discussed 
in detail in Sect.\ref{supertwistorsection}), and  a point in supertwistor space is nonlocally transformed to a  super light ray in  Minkowski superspace. Since the  space-time matrix is super-hermitian,  the  supertwistor variables satisfy the super null condition
\begin{equation}
Z_A^*Z^A=0,
\label{supertwisnul}
\end{equation}
where $Z_A^*$ denote the dual supertwistor defined by $Z_A^*=(Z_a^*,\xi_i^*)=(\pi_{\alpha}^*,{\omega^{\beta}}^*,\pi^*,\omega^*)$. 
Thus, the  present projective supertwistor  is  null, and carries (pseudo-)real $5|4$ degrees of freedom.
With given a super space-time point, the corresponding point in the supertwistor space is uniquely determined provided the lower components $\pi_A=(\pi_{\alpha},\pi)$ given.
 This indicates that  a  point in Minkowski superspace is nonlocally transformed to a supersphere $S^{2|2}$ in the projective supertwistor space. The super incidence relation is easily generalized to include ${N}$ flavor Grassmann odd coordinates:  
\begin{align}
&\omega^{\alpha}=ix^{\alpha\beta}\pi_{\beta}-i\theta_{\alpha i }\pi_i,
\nonumber\\
&\omega^i=-i\theta^{\alpha}_i\pi_{\alpha}+it\pi_i,
\label{morefermisuperinc}
\end{align}
where $i$ is the flavor index for  Grassmann odd coordinates, $i=1,2,\cdots, {N}$.
The corresponding supertwistor is $Z^A=(\omega^{\alpha},\pi_{\beta},\omega^i,\pi_i)$, and its dual is $Z_A^*=(\pi_{\alpha}^*,{\omega^{\beta}}^*,\pi_i^*,{\omega^i}^*)$. 
One may notice that the number of the fermion components in $Z^A$ is necessarily even, $2N$, due to the appearance of  pairs of  $\omega^i$ and $\pi_i$.

Here, we   comment differences between the present incidence relation (\ref{morefermisuperinc}) and   Ferber's  original relation \cite{FerberNucl.Phys.B132}:
\begin{align}
&\omega^{\alpha}=i(x^{\alpha\beta}+\frac{i}{2}{\theta_i^{\alpha}}^*\theta_i^{\beta})\pi_{\beta},\nonumber\\
&\omega^i=i\theta^{{\alpha}}_i\pi_{\alpha},
\label{ferbersinci}
\end{align}
where Grassmann coordinate index $i$ runs to arbitrary integer  $N$, and  $*$ represents the conventional complex conjugation. 
First of all, in Ferber's supertwistor, the supertwistors consist of  $(\omega^{\alpha}, \pi_{\alpha},\omega^i)$ and   fermionic counterparts of $\pi_{\alpha}$, namely $\pi_i$,  do not exist. The fermion components in the present  supertwistors are double compared to the original Ferber's  set-up, and this discrepancy becomes important in discussing the spin degrees of freedom in  quantum supertwistor theory (See Sect.\ref{secsupertwistorquantiz}). 
Next, 
 in Ferber's incidence relation (\ref{ferbersinci}), the space-time matrix is given by 
$x^{\alpha\beta}+\frac{i}{2}{\theta_i^{\alpha}}^*\theta_i^\beta$, and   is  $\it{not}$ hermitian due to the imaginary factor in front of  fermionic bilinears, while, in the present, the space-time matrix  is promoted to a super-hermitian matrix (\ref{superherx}). 
Besides,  in the present, the gauge freedom is the  bosonic  one (\ref{gaugesuperincidenceinx}) only, while in Ferber's,   fermionic gauge freedoms exist as well as  bosonic one:   
\begin{align}
&x^{\alpha\beta}\rightarrow x^{\alpha\beta}+a{\pi^{\alpha}}^*\pi^{\beta}-\frac{i}{2}(\beta_i{\theta_i^{\alpha}}^*\pi^{\beta}+\beta^*_{i}{\pi^{\alpha}}^*\theta^{\beta}_i),\nonumber\\
&\theta_i^{\alpha}\rightarrow \theta_i^{\alpha}+\beta_i\pi^{\alpha},
\end{align}
where $a$ is a Grassmann even real parameter and $\beta_i$ are  Grassmann odd complex  parameters.
The geometrical meaning of the bosonic gauge transformation is apparent as in the bosonic twistor:  a direction of a light ray in Minkowski space,  while the geometrical meaning of the fermionic gauge transformation is not clear.
Similarly,  a space-time point in Minkowski is  transformed to two-sphere (not supersphere) in the supertwistor space in Ferber's incidence relation. 

\section{massless particle in Minkowski super space-time}\label{supertwistorsection}

The massless particle set-up  provides a complementary physical 
approach to  the purely mathematical construction \cite{ptp7018}, and here,  such massless particle model for the geometrical supertwistor theory is explored. 
Hereafter, we consider Minkowski superspace $\mathcal{M}_{4|2}$  with the metric 
\begin{equation}
d\tau^2=\eta_{\mu\nu}dx^{\mu}dx^{\nu}+C_{\alpha\beta}d\theta^{\alpha}
d\theta^{\beta}.
\end{equation}
The  free particle action  in $\mathcal{M}_{4|2}$ is given by  
\begin{equation}
S=\frac{\mu}{2}\int d\tau ~(\eta_{\mu\nu}\dot{x}^{\mu}\dot{x}^{\nu}
+C_{\alpha\beta}\dot{\theta}^{\alpha}\dot{\theta}^{\beta}),
\label{actionfreeori}
\end{equation}
where $\mu$ denotes the mass of the particle and $\cdot$  the derivative about the  invariant  length $\tau$. 
Introducing the auxiliary variable $p_{\mu}$ and $p_{\alpha}$, (\ref{actionfreeori}) is rewritten as 
\begin{equation}
S=\int d\tau (\dot{x}^{\mu}p_{\mu}+\dot{\theta}^{\alpha}p_{\alpha}-\frac{1}{2\mu}p^{\mu}p_{\mu}-\frac{1}{2\mu}p^{\alpha}p_{\alpha}),
\end{equation}
where $x^{\mu}$, $\theta^{\alpha}$, $p^{\mu}$ and $p^{\alpha}$ are treated as 
independent variables.
We are interested in  the case of the massless particle in which  
$p^{\mu}$ and $p^{\alpha}$ satisfy the  super null condition: 
\begin{equation}
\eta_{\mu\nu}p^{\mu}p^{\nu}+C_{\alpha\beta}p^{\alpha}p^{\beta}=0.
\end{equation}
The super momenta,  $p^{\mu}$ and $p^{\alpha}$,  subject to the condition, can be  simply expressed as  the  bilinear forms of 3-component superspinor  $\pi_A=(\pi_1,\pi_2,\pi)^t$: 
\begin{equation}
p^0=\pi^{\ddagger}\pi,~~
p^{a}=2\pi^{\ddagger}l^{a}\pi,~~
p^{\alpha}=2\pi^{\ddagger}l^{\alpha}\pi.
\label{pandpalphahopfconst}
\end{equation}
$\pi_A$ are  the ``square root'' of the super null momenta, and 
  regarded as more fundamental variables than super momenta.
With use of the superspinor, the massless free  action becomes
\begin{align}
&S_0=\int d\tau \pi^*_A  \dot{x}^{AB} \pi_B\nonumber\\
&~~~=\int d\tau ( \dot{x}^{\alpha\beta} \pi^*_{\alpha} \pi_{\beta}+
\dot{\theta}^{\alpha} (\pi^*\pi_{\alpha}+C_{\alpha\beta}\pi^*_{\beta}
\pi) +\dot{x}^{33}\pi^*\pi),
\label{masslessSUSYparticleaction}
\end{align}
where $x^{AB}$ denotes the components of (\ref{explisupermatrix}).
$S_0$ is invariant under the global translation in the supertwistor space,
\begin{align}
&x^{AB}\rightarrow x^{AB}+c^{AB},\nonumber\\ 
&\pi_A \rightarrow \pi_A. 
\end{align}
In detail,  
\begin{equation}
x^{\alpha\beta}\rightarrow x^{\alpha\beta}+c^{\alpha\beta},~~\theta^{\alpha}\rightarrow \theta^{\alpha}+\gamma^{\alpha},
\end{equation}
where $c^{\alpha\beta}$ denote  Grassmann even constants and  
 $\gamma^{\alpha}$   Grassmann odd constants. 
From (\ref{masslessSUSYparticleaction}), the equations of motion for $x^{\mu}$ and $\theta_{\alpha}$ are derived as 
\begin{align}
&\frac{d}{d\tau}(\pi_{\alpha}^*\pi_{\beta}-\frac{1}{2}
\delta_{\alpha\beta}\pi^*\pi)=0,\nonumber\\
&\frac{d}{d\tau}(\pi^*\pi_{\alpha}+C_{\alpha\beta}\pi_{\beta}^*\pi)=0.
\end{align}
These provide 6 independent real equations, and suggest 
\begin{equation}
\dot{\pi}_A=0,
\label{derivativepisp}
\end{equation}
which  is consistent with the  assumption that the particle is free and hence carries  conserved momenta.
Similarly, the equations of motion of  $\pi_A$ are derived as 
\begin{equation}
\dot{x}^{\alpha\beta}\pi_{\beta}-C_{\alpha\beta}\dot{\theta}^{\beta}\pi=0,~~~\dot{x}^{33}\pi-\dot{\theta}^{\alpha}\pi_{\alpha}=0,\label{x33eom}
\end{equation}
or 
\begin{align}
&\dot{{x}}^{\alpha\beta}={\pi^{\alpha}}^*\pi^{\beta}-\frac{1}{2}
\delta^{\alpha\beta}\pi^*\pi,\nonumber\\
&\dot{\theta}^{\alpha}=-\frac{1}{2}{\pi_{\alpha}}^*\pi-\frac{1}{2}
\pi^*\pi^{\alpha},\nonumber\\
&\dot{x}^{33}=-\frac{1}{2}({\pi^1}^*\pi^1+{\pi^2}^*\pi^2-\pi^*\pi).
\label{classicalsosupert}
\end{align}
The right-hand-sides of (\ref{classicalsosupert}) are concisely represented by  the super momentum matrix $p$: 
\begin{equation}
p=2\eta_{\mu\nu}p^{\mu}l^{\nu}+2C_{\alpha\beta}p^{\alpha}l^{\beta}
=
\begin{pmatrix}
-p^0+p^3 & p^1-ip^2 &  -p^{\theta_2} \\
p^1+ip^2 & -p^0-p^3 &   p^{\theta_1} \\
-p^{\theta_1} &  -p^{\theta_2}       &   p^0 
\end{pmatrix}.
\end{equation}
From  (\ref{pandpalphahopfconst}), the components of $p$ are given by  
\begin{align}
&p^{\alpha\beta}=-2{\pi^{\alpha}}^*\pi^{\beta}+\delta^{\alpha\beta}\pi^*\pi,\nonumber\\
&p^{\alpha 3}={\pi^{\alpha}}^*\pi-\pi^*\pi_{\alpha},\nonumber
 \\
&p^{3 \alpha}= -\pi_{\alpha}^*\pi -\pi^*\pi^{\alpha},\nonumber\\ 
&p^{3 3}={\pi^1}^*\pi^1+{\pi^2}^*\pi^2-\pi^*\pi.
\label{componentsuperpmat}
\end{align}
Then,  (\ref{classicalsosupert}) is simply expressed as  
\begin{equation}
\dot{x}^{AB}=-\dot{a}(\tau) p^{AB},
\end{equation}
and  the solution is obtained as  
\begin{equation}
x^{AB}=x^{AB}_0-a(\tau)p_0^{AB},
\label{gaugefreexabmassless}
\end{equation}
where  we have used (\ref{derivativepisp}), and  $p_0^{AB}$ represent a constant super momentum matrix.
Substituting  (\ref{componentsuperpmat}) to (\ref{gaugefreexabmassless}),     one may find that the  gauge transformation   in the  super incidence relation (\ref{gaugesuperincidenceinx}) is  reproduced. 
Thus, the  massless particle formulation in Minkowski superspace presents a physical set-up for the geometrical supertwistor theory.

\section{Supertwistor action and  quantization }\label{secsupertwistorquantiz}

Generally, the gauge degree of freedom of the solution is  a consequence of that of  the action. Indeed, the massless superparticle action (\ref{masslessSUSYparticleaction}) is  invariant under the gauge transformation 
\begin{equation}
{x}^{AB}\rightarrow {x}^{AB}-a(\tau)p^{AB}.
\end{equation}
($\pi_A$ is a zero-mode of  $p^{AB}$:  $p^{AB}\pi_B=0$.)  Then, the  super space-time matrix $x^{AB}$ is a gauge dependent quantity, and the gauge invariant quantity is introduced as 
\begin{equation}
\omega^{A}=i{x}^{AB}\pi_{B}.
\label{supertwistorandspacetime}
\end{equation}
This is nothing but   the super incidence relation (\ref{superincimatr}).
Its (pseudo-)complex conjugation is given by 
\begin{equation}
{\omega^{A}}^*=-i\pi_B^* x^{BA}.
\label{supertwistorandspacetime*}
\end{equation}
Now, the super massless particle action (\ref{masslessSUSYparticleaction}) is concisely expressed  as  
\begin{equation}
S_0=-i\int d\tau ({\pi^A}^*\dot{\omega}_{A}+{\omega^B}^*\dot{\pi}_{B}),
\label{supertwistorfree2}
\end{equation}
and, with the supertwistor variables $Z^A=(\omega^A,\pi_A,\omega,\pi)$, 
 further simplified as   
\begin{equation}
S_0=-i\int d\tau Z_A^*\frac{d}{d\tau}Z^A,
\label{presentsupertwiact}
\end{equation}
where $Z^A$ and $Z_A^*$ represent the twistor and dual twistor variables
 subject to the constraint (\ref{supertwisnul}). 
Up to  total derivatives, the action (\ref{presentsupertwiact}) is  invariant under the global translation  in supertwistor space: 
\begin{equation}
Z^A\rightarrow Z^A+D^A
\end{equation}
with constant supertwistor $D^A$.
The supertwistor action (\ref{presentsupertwiact}) and the constraint (\ref{supertwisnul}) are  ``diagonalized'' by  recombination of the supertwistor variables:  
\begin{equation}
Z_D=\frac{1}{\sqrt{2}}
\begin{pmatrix}
1 & 0 & 1 & 0 & 0 & 0 \\
0 & 1 & 0 & 1 & 0 & 0 \\
1 & 0 & -1 & 0 & 0 & 0 \\
0 & 1 & 0 & -1 & 0 & 0 \\
0 & 0 & 0 & 0 & 1 & -1 \\
0 & 0 & 0 & 0 & -1 & 1
\end{pmatrix}
Z.
\end{equation}
With use of $Z_D$, the supertwistor norm is represented as 
\begin{equation}
Z_A^*Z^A={Z_D^1}^* Z_D^1 +{Z_D^2}^* Z_D^2 -{Z_D^3}^* Z_D^3 -{Z_D^4}^* Z_D^4 
+{\xi_D^1}^*\xi_D^1-{\xi_D^2}^*\xi_D^2,
\end{equation}
and thus the supertwistor space  has the metric: $diag(+,+,-,-,+,-)$.
The action and the null constraint are invariant under the  $SU(2,2|1,1)$ global transformation of the supertwistor variables.
Generally, with $N$-flavor fermionic coordinates,  the number of SUSY is    $\mathcal{N}=2N$, and  the global symmetry becomes  $SU(2,2|N,N)$.

Next, we discuss the quantization of  the supertwistor.
For simplicity, we consider one-flavor fermion case ($N=1$ then $\mathcal{N}=2$). With use of the bosonic and fermionic components of supertwistors, the  action (\ref{presentsupertwiact}) is  rewritten as 
\begin{equation}
S_0=-i\int d\tau (Z_{a}^*\frac{d}{d\tau}Z^{a}+\xi^*_i\frac{d}{d\tau}\xi^i), \label{supertwistoractionwithappvari}
\end{equation}
and the super null condition (\ref{supertwisnul}) becomes 
\begin{equation}
Z_{a}^*Z^{a}+\xi^*_i\xi^i=0.
\label{classicalconstraint}
\end{equation}
Apparently, (\ref{supertwistoractionwithappvari}) and (\ref{classicalconstraint})  are equal to what used  in  the original supertwistor theory \cite{ptp7018}, so  the quantization  reproduces  same physical contents as in the original supertwistor. 
We briefly explain the quantization procedure and results. 
From (\ref{supertwistoractionwithappvari}), the canonical conjugation of $Z^{a}$ is obtained as $-iZ_a^*$, and that of 
$\xi^{i}$ is $i\xi_i^*$.
Applying the canonical quantization condition to these variables  
\begin{equation}
[Z^a,Z^{*}_b]=-\delta^{a}_{~~b},~~\{\xi^i,\xi_j^*\}=\delta^{i}_{~~j},
\end{equation}
derivative expressions for  $Z^{*}_a$ and $\xi^*_i$  are obtained as   
\begin{equation}
Z^*_a=\frac{\partial}{\partial Z^a},~~~~\xi^*_i=\frac{\partial}{\partial \xi^i}.
\end{equation}
The super null condition (\ref{classicalconstraint}) is expressed as
 $\{Z^a,Z_a^*\}+[\xi^i,\xi_i^*]=0$ and  imposed to  
 the Hilbert space: 
\begin{equation}
(\{Z^a,Z_a^*\}+[\xi^i,\xi_i^*])|\Psi>=0.
\label{quantumconstraint}
\end{equation}
In the coordinate representation,  (\ref{quantumconstraint}) is rewritten as 
\begin{equation}
(Z^a\frac{\partial}{\partial Z^a}+\xi^i\frac{\partial}{\partial \xi^i}+1)\Psi=0,
\label{supernullcocoor}
\end{equation}
where $Z^a\frac{\partial}{\partial Z^a}$ is known as the Euler homogeneity operator. Then, $\Psi$ should be a homogeneous function of $Z^a$ and $\xi^i$, and 
the sum of the powers of $Z^a$ and $\xi^i$ should be   $-1$. 
Thus, in general, $\Psi$ is expressed as   
\begin{equation}
\Psi=t_{1/2}(Z^a)+t_0(Z^a)\xi^1+t'_0(Z^a)\xi^2+t_{-1/2}(Z^a)\xi^1\xi^2,
\end{equation}
where the expansion coefficients $t_s$ are called twistor functions, and 
 are given by 
\begin{equation}
t_{1/2}\!=\!h_{-1}(Z),~t_{0}\!=\!h_{-2}(Z),~t'_0\!=\!{h}'_{-2}(Z),~t_{-1/2}\!=\!h_{-3}(Z).
\label{lopsidedness}
\end{equation}
$h_{-n}(Z)$ represents a homogeneous function of $1/Z^n$.
In the twistor formulation \cite{TheCentralProgrammePenrose}, 
  the helicity is given by 
\begin{equation}
s=\frac{1}{2}Z_a^*Z^a,
\label{classicalheli}
\end{equation}
and  expressed as the  operator 
\begin{equation}
\hat{s}=\frac{1}{4}\{Z^a,Z_a^*\}=\frac{1}{2}Z^a\frac{\partial}{\partial Z^a}+1.\label{helicityop}
\end{equation}
The twistor functions $t_s$ are eigenfunctions of the helicity operator with  eigenvalue $s$, and are related by  supercharges  
\begin{equation}
Q_i^{~a}=\xi_i^*Z^a,~~~{Q_a^{~i}}^*=-Z_a^*\xi^i.
\end{equation}
$Q_i^{~a}$ and $Q_a^{~i}$ are  helicity $1/2$ and $-1/2$ operators, respectively: 
\begin{equation}
[\hat{s},Q_i^{~a}]=\frac{1}{2}Q_i^{~a},~~~[\hat{s},{Q_a^{~i}}^*]=-\frac{1}{2}{Q_a^{~i}}^*.
\end{equation}
Thus, the  number of SUSY (charges) is equal to that of the fermionic components of supertwistor $\xi^i$.  

It is straightforward to introduce $\mathcal{N}$ fermionic  components 
in supertwistors $\xi^i$ ($i=1,2,\cdots, \mathcal{N}$). 
In such $\mathcal{N}$-SUSY case, (\ref{supernullcocoor}) is generalized as 
\begin{equation}
(\hat{s}+\frac{1}{2}\xi^i\frac{\partial}{\partial\xi^i}-\frac{\mathcal{N}}{4})\Psi=0,
\end{equation}
where  $\frac{\partial}{\partial\xi^i} \xi_i=-\xi_i\frac{\partial}{ \partial \xi^i}+\mathcal{N}$ was used.
Since the operator $\xi^i\frac{\partial}{\partial \xi^i}$ can take  the eigenvalues $0,1,2\cdots,\mathcal{N}$,  the eigenvalues of the helicity operator  are distributed as  
\begin{equation}
s= -\frac{\mathcal{N}}{4},-\frac{\mathcal{N}}{4}+\frac{1}{2},-\frac{\mathcal{N}}{4}+1,\cdots,\frac{\mathcal{N}}{4}-\frac{1}{2},\frac{\mathcal{N}}{4}.\label{helicitydistri}
\end{equation}

Although the resultant quantum supertwistor  is superficially  equal to  that of the  original supertwistor \cite{ptp7018}, there are important differences.  
In the original supertwistor, the basic quantities are given by $x^{\mu}, p^{\mu}$  and $\theta^{\alpha}_i$ that amount to the complex coordinates: $y^{\mu}=\frac{1}{2}\sigma^{\mu}_{\alpha\beta}{\theta^{\alpha}_i}^* \theta^{\beta i}$ \cite{FerberNucl.Phys.B132}, while in the present, the basic quantities are $x^{\mu}, p^{\mu}, \theta^{\alpha}_i$ and $p^{\alpha}_i$ ($i=1,2,\cdots,N$), and complex space-time is  not introduced. In both approaches, the spin degrees of freedom are originated from the existence of the fermionic variables, since, from   the null supertwistor condition  (\ref{supertwisnul}),  the helicity $s$ (\ref{classicalheli}) is restated as   
\begin{equation}
s=-\frac{1}{2}\xi_i^*\xi^i.
\end{equation}
However, in the present geometrical formalism, the momentum space and the space-time are treated equivalently, and there always exist pairs of fermionic variables: $(p^{\alpha}_i, \theta^{\alpha}_i)$ or $(\pi_i,\omega^i)$.
Then, the helicity $s$ is expressed as  
\begin{equation}
s
=-\frac{1}{2}( \pi_i^*\omega^i + \omega_i^*\pi^i )
\end{equation}
with $\omega^i$ given by (\ref{morefermisuperinc}), 
and such fermion sets  amounts to even number SUSY $\mathcal{N}=2N$.  Meanwhile in the Ferber's original supertwistor, the helicity is given by  
\begin{equation}
s= -\frac{1}{2}\omega_i^*\omega^i
\end{equation}
with $\omega^i$ given by (\ref{ferbersinci}), and the number of SUSY is $\mathcal{N}=N$. 
Even number of SUSY is physically required to bring    integer of half-integer helicities (See (\ref{helicitydistri})), and it   has been fixed by hand in  the original supertwistor.
 Meanwhile in the geometrical construction, such condition is automatically satisfied because  $\mathcal{N}=2N$. 
Thus,  even  number of SUSY is necessarily incorporated in   the geometrical supertwistor. 
Besides,  in Ferber's approach the signatures of the fermionic space are not uniquely determined, while in the present  they  are unique:  $N$ for $+$  and $N$ for $-$.

\section{Relations to Superspin and SUSY Quantum Hall Effect}\label{relationstoquantumapin}

It is known that the (bosonic) Hopf map is a 
 mathematical background of   quantum mechanics of spin
\cite{quant-ph/0108137} and 
quantum Hall effect on two-sphere \cite{HaldanePRL51605}.
Here, we discuss how their structures are generalized and related  to 
the geometrical supertwistor  when the SUSY Hopf map is adopted.

\subsection{Relation to Superspin on Bloch supersphere}

In the context of spin quantum mechanics, the Hopf spinor is used to construct  a spin coherent state 
\begin{equation}
|\phi>=\phi_1|\uparrow>+\phi_2|\downarrow>,
\end{equation}
where $(\phi_1,\phi_2)^t$ is  the Hopf spinor (\ref{explihopfspi}) that specifies a point on 
 Bloch sphere by the Hopf map (\ref{explHopf}). 
It is well known that the $SU(2)$ spin mechanics is reformulated by  introducing  Schwinger bosons,  $a$ and $b$,  
 $|\uparrow>=a^{\dagger}|0>$, $|\downarrow>=b^{\dagger}|0>$.
In other words, the Hopf spinor is  a   coherent state representation of the Schwinger boson: 
\begin{equation}
<\phi|a>=\phi_1, ~~<\phi|b>=\phi_2.
\end{equation}
The spin magnitude corresponds to  half of the total number of  Schwinger bosons. For instance, to represent spin $1/2$, 
the  Schwinger boson operators satisfy  the constraint 
\begin{equation}
a^{\dagger}a+b^{\dagger}b=1.
\end{equation}
Meanwhile,  in the present, we have used  the super Hopf spinor which  
 contains  two Grassmann even  and one Grassmann odd components.
 Then, the  corresponding operators may be given by 
two bosonic operators $a$ and $b$, and one fermionic operator $f$:  
\begin{equation}
(\psi_1,\psi_2,\psi_0)\rightarrow (a,b,f).
\end{equation}
The normalization condition for the SUSY Hopf spinor is transformed to  the constraint of the operators:   
\begin{equation}
1=a^{\dagger}a+b^{\dagger}b+f^{\dagger}f, 
\end{equation}
which represents the superspin $1/2$.
Such formalism is known as the slave fermion formalism in condensed matter physics, where the fermionic operator is introduced to deal with the inequivalent condition 
\begin{equation}
a^{\dagger}a+b^{\dagger}b\le 1.
\end{equation}
Thus, in the slave fermion formalism,    spin $1/2$ and $0$  are treated simultaneously. 
With the super Hopf spinor (\ref{explicitsuperhopf}), the supersymmetric extension of the spin coherent state  is constructed as  
\begin{equation}
|\psi>=\psi_1|\uparrow>+\psi_2|\downarrow>+\psi_0|f>=\psi_1|a>+\psi_2|b>+\psi_0|f>,
\end{equation}
which is also known as the spin-hole state \cite{auerbachbook}.
 Thus, Bloch supersphere  is the hidden geometry behind the slave fermion formalism \footnote{The relevance to the slave fermion formalism was explained by the collaborators in \cite{AHQZ} to the author. }, and  
based on this observation, a supersymmetric  antiferromagnetic valence bond solid model was  constructed recently \cite{AHQZ}.

\subsection{Relation to SUSY Quantum Hall Effect}

Based on  the  SUSY Hopf map, a supersymmetric extension of the  quantum Hall effect is  constructed  in  \cite{hep-th/0409230,hep-th/0411137}, where  
the fermionic variables are  interpreted as spin degrees of freedom. 
In the present supertwistor model, the number of (minimal)  SUSY is $\mathcal{N}=2$, while  in the SUSY quantum Hall effect $\mathcal{N}=1$. 
This two-fold difference suggests the geometrical supertwistor may consist 
of two copies of the SUSY quantum Hall effect. 
Here, we pursue this heuristic observation. The supertwistor action (\ref{supertwistoractionwithappvari}) is rewritten as 
\begin{equation}
S= - i\int d\tau Z_+^{\ddagger}\frac{d}{d\tau} Z_+ + i\int d\tau Z_-^{\ddagger}\frac{d}{d\tau} Z_-,
\label{supertwistorqunatumhall}
\end{equation}
where $Z_+$ and $Z_-$ denote the diagonal supertwistors: $Z_+=(Z_D^1,Z_D^2,\xi_D^2)^t$ and $Z_-=(Z_D^3,Z_D^4,\xi_D^1)^t$, which  satisfy the super 
 null condition: $Z_+^{\ddagger}Z_+-Z^{\ddagger}_-Z_-=0$.
We focus on a ``slice'' of the null supertwistor space 
\begin{equation}
{Z_+}^{\ddagger}Z_+= {Z_-}^{\ddagger}Z_-=R^2,
\label{normlizaZ}
\end{equation}
with some constant $R$.   
From (\ref{normlizaZ}), the coordinates on supersphere are naturally defined as  
\begin{align}
&x^a=2Z_+^{\ddagger}l^a Z_+,~~\theta^{\alpha}=2Z_+^{\ddagger}l^{\alpha}Z_+,\nonumber\\
&y^a=2Z_-^{\ddagger}l^a Z_-,~~\vartheta^{\alpha}=2Z_i^{\ddagger}l^{\alpha}Z_-,
\end{align}
that satisfy the relation: $x^ax^a+C_{\alpha\beta}\theta^{\alpha}\theta^{\beta}=y^ay^a+C_{\alpha\beta}\vartheta^{\alpha}\vartheta^{\beta}=R^2$, and 
the SUSY monopole gauge fields are induced as 
\begin{align}
&-i Z_+^{\ddagger}{d} Z_+ = dx^a A_a+d\theta^{\alpha}A_{\alpha},\nonumber\\
&-i  Z_-^{\ddagger}{d} Z_- = dy^a A_a+d\vartheta^{\alpha}A_{\alpha}.
\end{align}
Then, the action (\ref{supertwistorqunatumhall}) becomes 
\begin{equation}
S=  \int d\tau \frac{d x^a}{d\tau}A_a +
\int d\tau \frac{d \theta_{\alpha}}{d\tau }A_{\alpha}
- \int d\tau \frac{d y^a}{d\tau }A_a -\int d\tau \frac{d \vartheta_{\alpha}}{d\tau }A_{\alpha},
\label{spinhalllikeaction}
\end{equation}
which  is formally equivalent to   two copies of one-particle action used in the  SUSY quantum Hall effect  by replacing invariant time  $\tau$ with  time $t$.
The opposite signs in front of the two copies suggest that the magnetic fields  are inversely aligned  in such two  copies.  
This   is something similar to  the spin Hall effect \cite{cond-mat/0308167},   where up-spin  and down-spin feel opposite effective magnetic fields.
Indeed,  the bosonic part of the action (\ref{spinhalllikeaction}) is  equal to what was used in the context of quantum spin Hall effect \cite{condmat0504147}.
Thus, in the slice of the supertwistor space, SUSY  spin Hall  analogous system is supposed to be realized. 

\section{Summary and Discussion}\label{summarysection}

We have geometrically constructed a supertwistor theory based on the 
 SUSY  Hopf map. 
The basic variables   are different from those of  Ferber's original supertwistor; fermionic  momenta  are newly introduced by  geometrical reasoning. The new super incidence relation is naturally derived based on the arguments of the celestial supersphere. 
The  super space-time matrix becomes  super-hermitian and  relates the  Minkowski superspace and the supertwistor space nonlocally in the sense: a point in  Minkowski superspace is transformed to a supersphere in the supertwistor space, and a point in the supertwistor space is transformed to a super light ray  in the Minkowski superspace. 
The quantum theory of the geometrical supertwistor reproduces  same physical contents as in  the original supertwistor theory, and besides, the present formalism has following remarkable properties. 
First of all, the 
space-time is not complexified to introduce the spin degrees of freedom. The space-time is promoted to a super-hermitian superspace and the (pseudo-)real fermion variables yield the origin of spin degrees of freedom.
Pairs of fermionic momenta and fermionic coordinates are  introduced, which necessarily amount to even number of SUSY to bring half integer or integer helicity states. We have also discussed relations to superspin quantum mechanics and the SUSY quantum Hall effect. 
Bloch supersphere is  the template geometry of the present model, and 
  provides the hidden geometry of the slave fermion formalism. With an appropriate choice of the slice of   supertwistor space, the SUSY  spin Hall analogous system is supposed to be realized. 
 Twistor theory shares  many analogous properties with quantum Hall effect, such as holomorphicity of  twistor functions and lowest Landau level functions, fuzzy geometry in space(-time) \cite{cond-mat/0211679,hep-th/0506120}. 
We would like report  detail analyses of their relations  elsewhere.  
The higher dimensional SUSY Hopf maps are proposed in \cite{hep-th/0611328}, and it is also interesting to see what geometrical supertwistor models come out based on  such  higher dimensional SUSY set-up.

\section*{ACKNOWLEDGMENTS}
I would like to thank Professor Hiroshi Kunitomo for valuable discussions on various aspects of twistor theory. I am also glad to thank Professor Shou-Cheng Zhang for giving me  stimulations for the study of  twistor theory.

\appendix

\section{Several definitions for supermatrix}\label{defofsumatapp}

We briefly summarize several definitions used in  super Lie group.
(For more detail, see \cite{dictionaryonsuperalgebras}.)
The super-conjugation acts to Grassmann odd quantities $\eta$ and $\xi$ as  
\begin{equation}
(\eta\xi)^*=\eta^*{\xi}^*, ~~(\eta^*)^*=-\eta.
\label{defofpseudo}
\end{equation}
With the supermatrix taking  the form of 
\begin{equation}
\begin{pmatrix}
A & B \\
C & D
\end{pmatrix},
\end{equation}
where $A$ and $D$ are Grassmann even component matrices, $B$ and $C$ are Grassmann odd component matrices, the super-adjoint $\ddagger$ is defined as 
\begin{equation}
\begin{pmatrix}
A & B \\
C & D
\end{pmatrix}^{\ddagger}
=
\begin{pmatrix}
A^{\dagger} & C^{\dagger} \\
-B^{\dagger} & D^{\dagger}
\end{pmatrix}.
\end{equation}
The supertrace is given by 
\begin{equation}
\text{str}\begin{pmatrix}
A & B \\
C & D
\end{pmatrix}
=\text{tr}A-\text{tr}D,
\end{equation}
and the superdeterminant is  
\begin{equation}
\text{sdet}\begin{pmatrix}
A & B \\
C & D
\end{pmatrix}
=\frac{\text{det}(A-BD^{-1}C)}{\text{det} D}=\frac{\text{det} A}{\text{det}(D-CA^{-1}B)}.
\end{equation}


\end{document}